\documentclass[singlespacing]{elsart}

\usepackage{graphicx}

\usepackage{amssymb}
\journal{}
\begin{document}

\begin{frontmatter}
\title{Analytic treatment of nuclear spin-lattice relaxation for diffusion in a cone model}

\author{A.E. Sitnitsky},
\ead{sitnitsky@mail.knc.ru}

\address{Institute of Biochemistry and Biophysics, P.O.B. 30, Kazan
420111, Russia. e-mail: sitnitsky@mail.knc.ru, Tel. 7-843-2319037, Fax. 7-843-2927347.}

\begin{abstract}
We consider nuclear spin-lattice relaxation rate resulted from a diffusion equation for rotational wobbling in a cone. We show that the widespread point of view that there are no analytical expressions for correlation functions for wobbling in a cone model is invalid and prove that nuclear spin-lattice relaxation in this model is exactly tractable and amenable to full analytical description.
The mechanism of relaxation is assumed to be due to dipole-dipole interaction of nuclear spins and is treated within the framework of the standard Bloemberger, Purcell, Pound - Solomon scheme. We consider the general case of arbitrary orientation of the cone axis relative the magnetic field. The BPP-Solomon scheme is shown to remain valid for systems with the distribution of the cone axes depending only on the tilt relative the magnetic field but otherwise being isotropic. We consider the case of random isotropic orientation of cone axes relative the magnetic field taking place in powders. Also we consider the cases of their predominant orientation along or opposite the magnetic field and that of their predominant orientation transverse to the magnetic field which may be relevant for, e.g., liquid crystals. Besides we treat in details the model case of the cone axis directed along the magnetic field. The latter provides direct comparison of the limiting case of our formulas with the textbook formulas for free isotropic rotational diffusion.
The dependence of the spin-lattice relaxation rate on the cone half-width yields results similar to those predicted by the model-free approach.

\end{abstract}

\begin{keyword}
NMR, spin-lattice relaxation, rotational diffusion, wobbling in a cone, model-free approach.
\end{keyword}
\end{frontmatter}

\section{Introduction}
Diffusometry and relaxometry  is a traditional and well developed branch of NMR \cite{Ric80}, \cite{Kim85}, \cite{Nus88}, \cite{Kim93}, \cite{Sta95}, \cite{Sta951},
\cite{Kim97}, \cite{Hal981}, \cite{Hal98}, \cite{Zav991}, \cite{Zav99}, \cite{Kle99}, \cite{Kle02}, \cite{Kim02}, \cite{Kim04}, \cite{Kne04}, \cite{Sit05}, \cite{Kne05}, \cite{Kne08}, \cite{Sit08}, \cite{Cal08}, \cite{Mag08}, \cite{Mag081}, \cite{Mag09}, \cite{Cal10}, \cite{Mei10}. It is successfully applied to many complex systems such as proteins and homopolypeptides \cite{Kim85}, \cite{Nus88}, \cite{Bue99}, \cite{Bue01}, \cite{Kor01}, \cite{Kor011}, \cite{Kim04}, \cite{Kor05}, \cite{Kor06}, \cite{Kor07}, \cite{God07}, \cite{God09}, \cite{God071}, \cite{Sun09}, tissues \cite{Kor02}, liquid crystals \cite{Hal98}, \cite{Leo04}, \cite{Kim04}, liquids in porous glass \cite{Sta951}, \cite{Kim04}, \cite{Sit05}, polymers \cite{Kim04}, etc. Purely empirical way to interpret relaxation behavior in complex systems is to introduce a distribution of the correlation times, e.g., Cole-Davidson or Cole-Cole ones (see e.g., \cite{Hal981}, \cite{Sul99}, \cite{Jar06}, \cite{Vog08} and refs. therein). Usually this is done within the framework of the so-called model-free approach \cite{Lip82}, \cite{Lip821}, \cite{Hal981}, \cite{Bue99}, \cite{Bue01}, \cite{Mod08}, \cite{Hal09}, \cite{Mei10}.

At the same time many important theoretical issues still remain open. For instance free isotropic rotational diffusion is well investigated in details \cite{Ab61}. However it is applicable to a limited number of cases. In practice one is encountered as a rule with some sort of restricted rotational diffusion, e.g., wobbling in a cone with half-width $\theta_0$ treated rigorously by Wang and Pecora \cite{Wan80}. The latter model is widely used for interpreting NMR relaxation data \cite{Ric80}, \cite{Gir05}, \cite{Jar06}, \cite{Lip801}, \cite{Lip811}. It has become a widespread point of view that there are no analytical expressions for correlation functions for wobbling in a cone model \cite{Lip801}, \cite{Lip811}, \cite{Gir05}.
For instance in \cite{Lip811} one can read: "although $< P_2\left(\hat \mu (0)\cdot \hat \mu (t)\right) >$ cannot be evaluated analytically within the diffusion in the cone model, ...". Analogous statement is reiterated in  \cite{Gir05}:"Although there is no analytical expression for $G_m (t)$ in the case of this diffusion in a cone model, ...". We do not agree with this point of view and the aim of the present paper is to show that nuclear spin-lattice relaxation in this model is exactly tractable and amenable to full analytic description. Here it is pertinent to recall a quotation from \cite{Lip811}:"To evaluate $G_m (t)$ exactly, one must solve time-dependent rotational diffusion equation subject to reflecting boundary conditions at $\theta_0$. Using this approach (Wang and Pecora \cite{Wan80}), one can express $G_m (t)$ as an infinite sum of exponentials, with amplitudes and time constants which are not closed-form functions of $\theta_0$".
Then the authors of \cite{Lip811} circumvent the problem by constructing a simple but accurate closed-form approximation to $G_m (t)$.
These results are sometimes referred to as quasi-exact \cite{Mar87} while the approximation is referred to as multi-exponential in contrast to the mono-exponential obtained in \cite{Lip801}.
In the present paper we show that direct and stringent way to tackle the problem is nevertheless quite feasible.

Thus in our opinion the quoted above assertion means not that statements like "there is no analytical expression for $G_m (t)$ in the case of the diffusion in a cone model" are valid but rather that exact analytic solution of the problem merely has not been obtained yet. In this regard the notion of analyticity should be refined. At the time the quotation from \cite{Lip811} was written only incomplete and erroneous Pal's tables for the values of $\nu_n^m$ (the roots of the derivative of the associated Legendre functions with respect to the degree for the cone half-width $\theta_0$) were available. This fact indeed made the expressions containing $\nu_n^m$ to be "not closed-form functions of $\theta_0$", the formulas to be practically intractable and as a consequence the resulting theory to be non-analytic. However the situation was dramatically improved shortly after the publication of \cite{Lip811}.
The appearance in $1986$ of the Bauer's tables for the values of $\nu_n^m$ as known functions of $\cos \theta_0$  \cite{Bau86} makes the problem of treating $\nu_n^m$ conceptually equivalent to that of any habitual function. One should consider $\nu_n^m$ as tabulated functions of $\theta_0$ just as for instance a trigonometric function of $\theta_0$. In this sense a formula containing $\nu_n^m (\theta_0)$ since $1986$ is as well analytic as that containing, e.g., $\sin (\theta_0)$. In our opinion the issue about analytic tractability of the correlation functions within the framework of the diffusion in a cone model has long been in need for revaluation.

In the present paper we precisely follow the instruction from the above quotation from \cite{Lip811}. We obtain analytic expressions for the exponentials from the infinite sum. Then with the help of a computer and Bauer's tables the problem of calculating the $G_m (t)$, the corresponding spectral densities and finally the spin-lattice relaxation rate becomes a routine though tedious procedure. We carry out these calculations for representative set of model parameters and various distributions of the cone axis relative the laboratory fixed frame (external magnetic field) and plot spin-lattice relaxation rates. The dependencies obtained exhibit rich variety of interesting non-monotonic behavior. Our rigorous quantum-mechanical treatment has a lucid classical analogy. The latter suggests some physical interpretation of the phenomenon. Also we find that our results are similar to those predicted by the model-free approach.
Thus in our opinion the problem of calculating the correlation functions and spin-lattice relaxation rates within the framework of diffusion in a cone model gets full analytic solution. The same is true for spin-spin relaxation rate though this value is not addressed in the present paper to save room.

The approach is discussed mainly for practically important case of an isolated two-spin system comprising a hetero-nuclear pair of non-identical spins, e.g., $^{15}N - H$ in protein backbone or $^{13}C - H$ in protein side chains. However the results for homo-nuclear spin pair are also presented for the sake of completeness. For wobbling in a cone we can extend the cone half-width $\theta_0$ up to the limit of isotropic rotation $\theta_0\rightarrow\pi$ to verify the coincidence of the results obtained with the known formulas.
Only for the artificial potential of wobbling in a cone the equation for restricted rotational diffusion can be solved in a stringent way without any approximation. For no other non-trivial model such exact treatment is possible. Thus the results of rigorous treatment of the rotational diffusion equation for wobbling in a cone can serve as a touchstone for approximations by necessity invoked to in the case of more realistic potentials, such as, e.g., a harmonic one. In this regard the wobbling in a cone model is distinguished in the universe of models for restricted rotational motion. All other models require severe approximations and so long as the wobbling in a cone model is treated approximately it is merely one among many others. But as soon as this model is treated rigorously it takes an outstanding position of the exactly solvable one. Namely this fact motivates our attempt to revisit the investigation of the thirty years old problem of applying the Wang-Pecora model to NMR.

We discuss the possible manifestation of the deviations of rigorous results for the spin-lattice relaxation rate from those of the approximate approach \cite{Lip82}, \cite{Lip811} in the experiment. The requirements for the possibility of observing the above mentioned non-monotonic behavior in the experiment are found to be rather severe. Thus we anticipate that it can manifest itself provided it is sought purposefully and special concern is paid to the requirements. In these particular experimental situations the rigorous quantum-mechanical treatment developed in the present paper is expected to be more accurate compared with the approximate one of \cite{Lip82}, \cite{Lip811}. By present no such experiments have been carried out. The rigorous treatment of the diffusion in a cone model enables one to reveal very subtle peculiarities in the behavior of the spin-lattice relaxation rate. That is why to verify them reliably in the experiment its special settings up should be deliberately devised. The advantage of the present approach compared with  the approximate one lies mainly in the fact that the resulting theoretical description is completely coherent, i.e., results of the quantum-mechanical treatment agrees with those of its classical analogy and with those of \cite{Lip82}, \cite{Lip811}. In our opinion such full picture enables one to gain more penetrating into the diffusion in a cone model and its possibilities to describe nuclear spin-lattice relaxation.

The paper is organized as follows. In Sec.2 the diffusion equation for restricted rotational wobbling in a cone is used to derive the joint probability density function. In Sec.3 the latter is used to obtain the spectral densities of correlation functions for dipole-dipole interaction within the framework of the standard Bloemberger, Purcell, Pound (BPP) - Solomon scheme.  Sec. 4-5 deal with the particular case that the cone axis is directed along the magnetic field. In Sec.4 the spin-lattice relaxation rate for hetero-nuclear spin pair is obtained. In Sec.5 that for homo-nuclear spin pair is obtained. In Sec.6 the general case of arbitrary orientation of the cone axis relative the magnetic field is considered. These results are applied to the case of isotropic random orientation (unweighted average) of cone axes relative the laboratory fixed frame. Also a model example of predominant orientation of cone axes along or opposite the magnetic field and that of their predominant orientation transverse to the magnetic field are considered. In Sec.7 the results are discussed and the conclusions are summarized. In Appendix A some known mathematical formulas are collected for the convenience of the reader. In Appendix B some technical details of calculations are presented. In Appendix C the classical analogy of the quantum-mechanical treatment is considered.

\section{Rotational diffusion in a cone}
We choose a laboratory fixed frame so that its $z$ axis of the Cartesian frame  ${x,y,z}$ is directed along the constant magnetic field. The random functions $F^{(q)}$ of relative positions of two spins specified below (see (\ref{eq14})) are defined in the corresponding spherical frame ${\theta, \phi}$ given by the polar angle $\theta$ (counted from the $z$ axis) and azimuthal one $\phi$. We consider a general case that the cone axis is tilted at an angle $\psi$ relative the magnetic field. We direct the $z'$ axis of the dashed Cartesian frame ${x',y',z'}$ along the cone axis. The correlation function for wobbling in this cone we define in the corresponding spherical frame ${\theta', \phi'}$ given by the polar angle $\theta'$ (counted from the $z'$ axis) and azimuthal one $\phi'$.
Following \cite{Wan80} we consider a rod with the orientation specified by a unit
vector $\hat u$ directed along its axis with spherical polar coordinates
$\Omega' =(\theta', \phi')$. In accordance with \cite{Ab61} we following Debye assume that the rotation of the rod can be considered as that of the hard sphere with radius $a$ ($a$ is the length of the rod) in a medium with viscosity $\eta$.
For the ordinary diffusion in a cone model the rod is allowed to diffuse freely
within an empty cone with a maximum polar angle
$\theta'=\theta_0$. The symmetry axis of the
cone is taken to be the $z'$ axis. For the diffusion in a
cone model, the polar angle is restricted ($0 \leq \theta' \leq \theta_0$) but
the azimuthal angle is not ($0 \leq \phi' \leq 2\pi$).

Our aim is to consider ordinary diffusion for rotational motion in a cone.
The probability density for finding the rod oriented in
$\hat u$ at time $t$, i. e., $\Psi(\hat u, t)$, obeys the DE for rotational motion
\begin{equation}
\label{eq1} \frac{\partial \Psi(\hat u, t)}{\partial t}=D\Delta \Psi(\hat u, t)
\end{equation}
where  $D$ is the  diffusion coefficient (DC) for rotation and $\Delta$ is the angular part of the Laplace operator in polar spherical coordinates
\begin{equation}
\label{eq2} \Delta=\frac{1}{\sin^2 \theta'}\left[\sin \theta'\frac{\partial}{\partial \theta'}
\left(\sin \theta' \frac{\partial}{\partial \theta'}\right)+\frac{\partial^2}{\partial \phi'^2}\right]
\end{equation}
The DC for rotation  $D$ has the dimension $cm^2/s$ and is given by the Stokes formula
\begin{equation}
\label{eq3}  D =\frac{k_B T}{8\pi a^3\eta}
\end{equation}
where $k_B$ is the Boltzman constant and $T$ is the temperature.

Following \cite{Wan80} we write the solution of (\ref{eq1}) as follows
\[
 \Psi(\hat u, t)=
\]
\begin{equation}
\label{eq4} \sum_{n=1}^{\infty}\sum_{m=-\infty}^{\infty}
\exp\left[-\nu_n^m(\nu_n^m+1)D\mid t \mid \right]
Y_{\nu_n^m}^{(m)\ \ast}\left(\Omega'(0)\right)Y_{\nu_n^m}^{(m)}\left(\Omega'(t)\right)
\end{equation}
where $Y_{\nu_n^m}^{(m)}\left(\Omega'\right)$ is the generalized spherical harmonics of degree $\nu_n^m$ \cite{Wan80}, the symbol $\ast$ indicates the complex conjugate and the values of $\nu_n^m$ are determined by the boundary conditions on $\theta'$ defined by our diffusion in a cone model. The boundary condition
says that there is no net change of the probability density
at the boundary of the cone, i.e.,
\begin{equation}
\label{eq5}  \frac{\partial \Psi(\hat u, t)}{\partial \theta'}
\left| {\begin{array}{l}
  \\
\theta'=\theta_0\\
 \end{array}}\right. =0
\end{equation}
The values $\nu_n^m$ are known functions of $\cos \theta_0$ \cite{Wan80},  \cite{Bau86}.
They satisfy the requirement $\nu_n^{-m}=\nu_n^m$ .
The index $n$ is defined such that $\nu_1^m < \nu_2^m < \nu_3^m <...$ for any $m$.
The detailed calculations of $\nu_n^m$ are presented in the tables \cite{Bau86}.
The values of $\nu_n^m$ for $n > 1$ increase sharply with the decrease of the confining volume.

The solution (\ref{eq4}) is subjected to the initial condition
\begin{equation}
\label{eq6}  \Psi(\hat u, 0)=\delta \left(\Omega'-\Omega'(0)\right)=
\delta\left(\cos \theta' -\cos \theta'(0)\right)\delta\left(\phi'-\phi'(0)\right)
\end{equation}
The joint probability of finding the rod with orientation
$\hat u(0)$ in solid angle $d\Omega'(0)$ at time $0$ and $\hat u(t)$ in  $d\Omega'(t)$ at
time $t$ can be written as
\[
p\left( \Omega'(t),t; \Omega'(0),0 \right)=\frac{1}{2\pi (1-\cos \theta_0)}\times
\]
\begin{equation}
\label{eq7}
\sum_{n=1}^{\infty}\sum_{m=-\infty}^{\infty}
\exp \left[-\nu_n^m(\nu_n^m+1)D\mid t \mid\right]
Y_{\nu_n^m}^{(m)\ \ast}\left(\Omega'(0)\right)Y_{\nu_n^m}^{(m)}\left(\Omega'(t)\right)
\end{equation}
The latter satisfies the normalization condition
\begin{equation}
\label{eq8} \int\limits_{cone} \int\limits_{cone} p\left( \Omega'(t),t; \Omega'(0),0 \right)d\Omega'(t)d\Omega'(0)=1
\end{equation}
where the angular integrals are taken only over the volume
of the cone.

\section{NMR framework for rotational diffusion}
At the beginning of this Sec. we recall the main facts from the general theory of spin-lattice relaxation by dipole-dipole interaction suggested by BPP and developed by Solomon \cite{Sol55}. The BPP-Solomon scheme is substantiated by more stringent Redfield's theory  (see \cite{Ab61} for detailed presentation). This scheme is developed in the frame whose $z$ axis is directed along the constant magnetic field. For the case of identical spins $I$ the contribution
to the spin-lattice relaxation rate constant due to rotational
diffusion with the spectral density at a Larmor frequency $\omega
_{L}$ of the correlation function for spherical harmonics has the
form (see VIII.76 in \cite{Ab61})
\begin{equation}
\label{eq9} \left(1/T_{1}\right)_{rotat}=\frac{3\gamma^4\hbar^2I(I+1)}{2}
\left \{ J^{(1)}(\omega _{L})+ J^{(2)}(2\omega _{L})\right \}
\end{equation}
where $\gamma$ is the gyromagnetic ratio of the nucleus, $I$ is
their spin and $\hbar$ is the Planck constant. For non-identical spins $I$ and $S$ we have four equations (see VIII.88 in \cite{Ab61})
\[
\left(1/T_{1}^{II}\right)_{rotat}=\gamma_I^2\gamma_S^2 \hbar^2 S(S+1)\times
\]
\[
\left \{\frac{1}{12}J^{(0)}\left(\omega_{L}^I-\omega_{L}^S\right)+
\frac{3}{2}J^{(1)}\left(\omega_{L}^I\right)+
\frac{3}{4}J^{(2)}\left(\omega_{L}^I+\omega_{L}^S\right)\right \}
\]
and
\[
\left(1/T_{1}^{IS}\right)_{rotat}=\gamma_I^2 \gamma_S^2 \hbar^2 I(I+1)\times
\]
\begin{equation}
\label{eq10} \left \{-\frac{1}{12}J^{(0)}\left(\omega_{L}^I-\omega_{L}^S\right)+ \frac{3}{4}J^{(2)}\left(\omega_{L}^I+\omega_{L}^S\right)\right \}
\end{equation}
Here only two equations are written out explicitly because the other two can be obtained from them by mere changing of indexes \cite{Ab61}.
To find the spectral densities $J^{(0)}(\omega)$, $J^{(1)}(\omega)$ and $J^{(2)}(\omega)$ we need to know the correlation functions $G^{(i)}(t)$ where $i=0, 1, 2\ $.

As was stressed in the previous Sec. we set $\theta'$ and $\phi'$ to be polar angles defining the direction of the axis connecting protons and $\Psi(\theta', \phi', t)=\Psi \left(\Omega', t\right)$ to be the probability of the orientation of this axis in the direction $\Omega'$ at time $t$. In the general case the axis of the cone can be tilted relative the magnetic field ($z$ axis of the laboratory fixed frame) at an arbitrary angle $\psi$. That is why at application to a realistic system the correlation function of internal motion in the cone $<F^{(i)}(0)F^{(i)}(t)>_{internal}$ has to be averaged over the orientations of cone axes with some overall distribution of the angles $f(\psi)$ characterizing the system of interest. That is the correlation function $G^{(i)}(t)$ whose spectral densities are to be substituted into (\ref{eq9}) or (\ref{eq10}) has the form
\[
G^{(i)}(t)=<<F^{(i)}(0)F^{(i)}(t)>_{internal}>_{overall}=
\]
\begin{equation}
\label{eq11}\frac{1}{2} \int \limits_{0}^{\pi} d\psi\ \sin \psi\ f(\psi )<F^{(i)}(0)F^{(i)}(t)>_{internal}
\end{equation}
To put it differently we assume that the distribution function $f(\psi, \lambda, \omega)$ (where $\psi, \lambda, \omega$ are Euler angles for rotation of the dashed Cartesian frame ${x',y',z'}$ relative the laboratory fixed one ${x,y,z}$) depends only on the Euler angle $\psi$, i.e., $f(\psi, \lambda, \omega)\equiv f(\psi)$.
It will be shown in Sec.6 that only in this case the overall averaging provides the absence of cross-correlational functions with $q \not= q'$, i.e.,
\begin{equation}
\label{eq12} << F^{(q)}(0)F^{(q^{\prime})}(t)>_{internal}>_{overall}=\delta_{qq^{\prime}}G^{(q)}(t)
\end{equation}
where $\delta_{nm}$ is the Kronecker symbol. The latter requirement is crucial for the validity of the BPP-Solomon scheme \cite{Ab61}.
Further we consider four particular cases.

a). The cone axis is directed along the magnetic field for all cones in the system. It means that $f(\psi)=\delta (\psi)$ where $\delta (x)$ is a Dirac $\delta-$function (see Appendix B for technical details). This case is of little practical significance. However it provides direct comparison of the limiting case of our results with the textbook formulas from \cite{Ab61}. Thus it serves as a test for the validity of the present approach from the theoretical side. Besides the formulas obtained in this case without superfluous complexities are further used in more involved cases as building blocks. That is why we denote the correlation functions $G^{(i)}(t)$ for this case as basic ones $g^{(i)}(t)\equiv G^{(i)}(t)_{f(\psi)=\delta (\psi)}$. This case is considered in details in Sec.4-Sec.5.

b). For the particular case of random isotropic distribution (unweighted average) of cone axes relative the magnetic field  we have $f(\psi)=1$. This case is considered in Sec.6.

c). As an example of the case for the cone axes to be predominantly oriented along or opposite the magnetic field we consider the model function $f(\psi)=3\cos^2 \psi$. This case is considered in Sec.6.

d). As an example of the case for the cone axes to be predominantly oriented transverse to the magnetic field we consider the model function $f(\psi)=3/2\sin^2 \psi$. This case is considered in Sec.6.

From now and up to the end of Sec.5 we consider the case a)., i.e., $f(\psi)=\delta (\psi)$ where $\delta (x)$ is a Dirac $\delta-$function.
In this case each cone axis is directed along the magnetic field and we need not distinguish the dashed Cartesian frame ${x',y',z'}$ from the laboratory fixed one ${x,y,z}$. To retain the uniformity of designations for correlation function of wobbling in a cone we further use the dashed Cartesian frame and correspondingly the dashed spherical frame ${\theta', \phi'}$.
We start from the general expression for the correlation functions of arbitrary order (see VIII.13 in  \cite{Ab61}) that in our case takes the form
\[
g^{(i)}(t)\equiv G^{(i)}(t)_{f(\psi)=\delta (\psi)}= <F^{(i)}(0)F^{(i)}(t)>_{internal}=
\]
\begin{equation}
\label{eq13} \int\limits_{cone} \int\limits_{cone} F^{(i)\ \ast}\left(\Omega'(t)\right) F^{(i)}\left(\Omega'(0)\right)
p\left( \Omega'(t),t; \Omega'(0),0 \right)d\Omega'(t)d\Omega'(0)
\end{equation}
where $i=0, 1, 2$ and the angular integrals are taken only over the volume
of the cone.
We need the correlation functions $g^{(0)}(t)$, $g^{(1)}(t)$ and $g^{(2)}(t)$ in order to substitute their spectral densities in (\ref{eq9}) or (\ref{eq10}). For our case of dipole-dipole interaction of two spins separated by the distance $b$ they are defined by random functions $F^{(0)}$, $F^{(1)}$ and $F^{(2)}$ \cite{Ab61} whose relationship with associated Legendre functions $P_{2}^{(q)}\left(\cos \theta\right)$ is known (see, e.g., Appendix c. in \cite{Lan74})
\[
 F^{(0)}\left(\Omega\right)=\frac{1-3\cos^2 \theta }{b^3}=
-\frac{2}{b^3}P_{2}^{(0)}\left(\cos \theta\right)
\]
\[
 F^{(1)}\left(\Omega\right)=\frac{\sin \theta \cos \theta \exp (i\phi)}{b^3}=
\frac{1}{3b^3}P_{2}^{(1)}\left(\cos \theta\right)\exp (i\phi)
\]
\begin{equation}
\label{eq14} F^{(2)}\left(\Omega\right)=\frac{\sin^2 \theta  \exp (2i\phi)}{b^3}=
\frac{1}{3b^3}P_{2}^{(2)}\left(\cos \theta\right)\exp (i2\phi)
\end{equation}
We stress once more that in our particular case $f(\psi)=\delta (\psi)$ the $F^{(i)}\left(\Omega\right)$ in the laboratory fixed frame are identical to $F^{(i)}\left(\Omega'\right)$ in the dashed (cone-related) frame to be substituted in (\ref{eq13}).

Now we have to substitute (\ref{eq7}) and (\ref{eq14}) into (\ref{eq13}). We denote
\begin{equation}
\label{eq15} \mu=\cos \theta'
\end{equation}
so that $\mu_0=\cos \theta_0$ and introduce the associated Legendre functions $P_{\nu_n^m}^{(m)}\left(\mu\right)$
\begin{equation}
\label{eq16} Y_{\nu_n^m}^{(m)}\left(\Omega'\right)=\sqrt{\frac{1}{2\pi H_n^{(m)}}}
\exp (im\phi')P_{\nu_n^m}^{(m)}\left(\mu\right)
\end{equation}
which satisfy the orthogonality properties
\begin{equation}
\label{eq17} \int \limits_{\mu_0}^1 d\mu\  P_{\nu_{n_1}^m}^{(m)}\left(\mu\right)
P_{\nu_{n_2}^m}^{(m)}\left(\mu\right)=H_{n_1}^{(m)}\delta_{n_1, n_2}
\end{equation}
where $\delta_{n, m}$ is the Kronecker symbol ($\delta_{n, m}=1$ if $n=m$ and $\delta_{n, m}=0$ otherwise).

Making use of 1.12.1.12 and 1.12.1.9 from \cite{Pru03} (see Appendix A) respectively we obtain after straightforward calculations
\[
g^{(1)}(t)=\frac{1+\mu_0}{b^6}
\sum_{n=1}^{\infty}
\exp\left[-\nu_n^1(\nu_n^1+1)D\mid t \mid \right]\frac{1}{H_n^{(1)}}\times
\]
\begin{equation}
\label{eq18} \frac{1}{(\nu_n^1+3)^2(\nu_n^1-2)^2}
\left\lbrace \left[(\nu_n^1+3)\mu_0^2-1 \right]P_{\nu_{n}^1}^{(1)}\left(\mu_0\right)-
\nu_n^1 \mu_0 P_{\nu_{n}^1+1}^{(1)}\left(\mu_0\right)\right\rbrace^2
\end{equation}
and
\[
g^{(2)}(t)=\frac{(1-\mu_0)^2(1+\mu_0)^3}{b^6}
\sum_{n=1}^{\infty}
\exp\left[-\nu_n^2(\nu_n^2+1)D\mid t \mid \right]\frac{1}{H_n^{(2)}}\times
\]
\begin{equation}
\label{eq19} \frac{1}{(\nu_n^2+3)^2(\nu_n^2-2)^2}
\left\lbrace P_{\nu_{n}^2}^{(3)}\left(\mu_0\right)\right\rbrace^2
\end{equation}

The calculation of $G^{(0)}(t)$ requires the formula from \cite{Wan80}
\[
K_n^0=\int \limits_{\mu_0}^1 d\mu\  (3\mu^2-1)P_{\nu_{n}^0}^{(0)}=
4z_0 \Biggl [ \left(1-6z_0+6z_0^2\right)F\left(-\nu_{n}^0,\nu_{n}^0+1;2;z_0\right)+
\]
\begin{equation}
\label{eq20}
3z_0\left(1-2z_0\right)F\left(-\nu_{n}^0,\nu_{n}^0+1;3;z_0\right)+
2z_0^2F\left(-\nu_{n}^0,\nu_{n}^0+1;4;z_0\right)\Biggr ]
\end{equation}
where
\begin{equation}
\label{eq21} z_0=\frac{1-\mu_0}{2}
\end{equation}
and $F(a,b;c;x)$ is a hypergeometric function. Making use of (\ref{eq20}) we obtain
\begin{equation}
\label{eq22} g^{(0)}(t)=\frac{1}{(1-\mu_0)b^6}
\sum_{n=1}^{\infty}
\exp\left[-\nu_n^0(\nu_n^0+1)D\mid t \mid \right]\frac{1}{H_n^{(0)}}
\left\lbrace K_n^0\right\rbrace^2
\end{equation}

We denote
\begin{equation}
\label{eq23} \tau^{(m)}_n=\frac{1}{\nu_n^m(\nu_n^m+1)D}
\end{equation}
where $m=0,1,2$. We obtain
for the basic spectral densities $j^{(0)}(\omega)$, $j^{(1)}(\omega)$ and $j^{(2)}(\omega)$ of $g^{(0)}(t)$, $g^{(1)}(t)$ and $g^{(2)}(t)$ respectively
\begin{equation}
\label{eq24} j^{(0)}(\omega )=\frac{2}{(1-\mu_0)b^6}\sum_{n=1}^{\infty}\frac{\tau^{(0)}_n}{1+\left (\omega \tau^{(0)}_n\right )^{2}}\frac{1}{H_n^{(0)}}
\left\lbrace K_n^0\right\rbrace^2
\end{equation}
and
\[
j^{(1)}(\omega )=\frac{2}{b^6}
\sum_{n=1}^{\infty}\frac{\tau^{(1)}_n}{1+ \left (\omega \tau^{(1)}_n\right )^{2}}\frac{1}{H_n^{(1)}}\times
\]
\begin{equation}
\label{eq25} \frac{(1+\mu_0)}{(\nu_n^1+3)^2(\nu_n^1-2)^2}
\left\lbrace \left[(\nu_n^1+3)\mu_0^2-1 \right]P_{\nu_{n}^1}^{(1)}\left(\mu_0\right)-
\nu_n^1 \mu_0 P_{\nu_{n}^1+1}^{(1)}\left(\mu_0\right)\right\rbrace^2
\end{equation}
and
\begin{equation}
\label{eq26} j^{(2)}(\omega )=\frac{2}{b^6}
\sum_{n=1}^{\infty}\frac{\tau^{(2)}_n}{1+\left (\omega \tau^{(2)}_n\right )^{2}}\frac{(1-\mu_0)^2(1+\mu_0)^3}{H_n^{(2)}(\nu_n^2+3)^2(\nu_n^2-2)^2}
\left\lbrace P_{\nu_{n}^2}^{(3)}\left(\mu_0\right)\right\rbrace^2
\end{equation}

These spectral densities enable us to calculate any of the spin-lattice relaxation rates (\ref{eq9})-(\ref{eq10}) for the particular case that the cone axis is directed along the magnetic field.

We denote the correlation time $\tau$ and the rotational correlation time $\tau_{rotat}$
\begin{equation}
\label{eq27} \tau =\frac{8\pi a^3\eta}{k_B T}\ \ \ \ \ \ \ \ \ \ \ \ \ \ \ \ \ \ \ \ \ \ \ \ \ \ \ \
\tau_{rotat} =\frac{4\pi a^3\eta}{3k_B T}=\frac{\tau}{6}
\end{equation}

\section{Spin-lattice relaxation rate for nuclear pair with non-identical spins}
For the particular case that the cone axis is directed along the magnetic field we have for the spectral densities to be substituted in (\ref{eq9})-(\ref{eq10}) are: $J^{(0)}(\omega )\equiv j^{(0)}(\omega )$,  $J^{(1)}(\omega )\equiv j^{(1)}(\omega )$ and $J^{(2)}(\omega )\equiv j^{(2)}(\omega )$ where $j^{(0)}(\omega )$, $j^{(1)}(\omega )$ and $j^{(2)}(\omega )$ are given by (\ref{eq24})-(\ref{eq26}).
It is worthy to note that if we identify $S$ with, e.g., $^{15}N$ from a nuclear pair of non-identical spins $^{15}N - H$
then we actually need only the formula for $\left(1/T_{1}^{II}\right)_{rotat}$ (see (\ref{eq10})) for the analysis of experimental data. That is why further we restrict ourselves only with explicit writing out the formula for this quantity. The substitution of  (\ref{eq24})-(\ref{eq26}) into (\ref{eq10}) yields
\[
\left( \frac{b^6}{\tau\gamma_I^2\gamma_S^2 \hbar^2 S(S+1)}\right)
\left(\frac{1}{T_{1}^{II}}\right)_{rotat}=\sum_{n=1}^{\infty}\Biggl \{\frac{\left\lbrace K_n^0\right\rbrace^2}{6(1-\mu_0)H_n^{(0)}\nu_n^0(\nu_n^0+1)}
\times
\]
\[
\left [1+ \frac{\left (\left (1-\gamma_S/\gamma_I\right )\omega_{L}^I \tau \right )^{2}}{\left[\nu_n^0(\nu_n^0+1)\right]^{2}}\right]^{-1}+
\frac{\left[\nu_n^1(\nu_n^1+1)\right]^{-1}}{1+\left (\omega_{L}^I \tau\right )^{2}\left[\nu_n^1(\nu_n^1+1)\right]^{-2}}\times
\]
\[
\frac{3(1+\mu_0)}{H_n^{(1)}(\nu_n^1+3)^2(\nu_n^1-2)^2}
\left\lbrace \left[(\nu_n^1+3)\mu_0^2-1 \right]P_{\nu_{n}^1}^{(1)}\left(\mu_0\right)-
\nu_n^1 \mu_0 P_{\nu_{n}^1+1}^{(1)}\left(\mu_0\right)\right\rbrace^2+
\]
\[
\left [1+ \frac{\left (\left (1+\gamma_S/\gamma_I\right )\omega_{L}^I \tau \right )^{2}}{\left[\nu_n^2(\nu_n^2+1)\right]^{2}}\right ]^{-1}\times
\]
\begin{equation}
\label{eq28} \frac{3(1-\mu_0)^2(1+\mu_0)^3}{2H_n^{(2)}(\nu_n^2+3)^2(\nu_n^2-2)^2\nu_n^2(\nu_n^2+1)}
\left\lbrace P_{\nu_{n}^2}^{(3)}\left(\mu_0\right)\right\rbrace^2\Biggr \}
\end{equation}

This formula describes the spin-lattice relaxation rate from restricted rotational diffusion in a cone for the particular case of the cone axis to be directed along the magnetic field. The series in this formula is well convergent. That is why in practice it is sufficient to restrict oneself only by several initial terms in it.

\section{Spin-lattice relaxation rate for nuclear pair with identical spins}
For the case of identical spins the substitution of (\ref{eq24})-(\ref{eq26}) into (\ref{eq9}) yields
\[
\left( \frac{b^6}{\tau\gamma^4 \hbar^2 I(I+1)}\right)
\left(\frac{1}{T_{1}}\right)_{rotat}=3\sum_{n=1}^{\infty}\Biggl \{\frac{1}{\nu_n^1(\nu_n^1+1)}
\frac{1}{1+ \left (\omega_{L} \tau\right )^{2}\left[\nu_n^1(\nu_n^1+1)\right]^{-2}}\times
\]
\[
\frac{(1+\mu_0)}{H_n^{(1)}(\nu_n^1+3)^2(\nu_n^1-2)^2}
\left\lbrace \left[(\nu_n^1+3)\mu_0^2-1 \right]P_{\nu_{n}^1}^{(1)}\left(\mu_0\right)-
\nu_n^1 \mu_0 P_{\nu_{n}^1+1}^{(1)}\left(\mu_0\right)\right\rbrace^2+
\]
\[
\frac{1}{1+ \left (2\omega_{L} \tau \right )^{2}\left[\nu_n^2(\nu_n^2+1)\right]^{-2}}\times
\]
\begin{equation}
\label{eq29} \frac{(1-\mu_0)^2(1+\mu_0)^3}{H_n^{(2)}(\nu_n^2+3)^2(\nu_n^2-2)^2\nu_n^2(\nu_n^2+1)}
\left\lbrace P_{\nu_{n}^2}^{(3)}\left(\mu_0\right)\right\rbrace^2\Biggr \}
\end{equation}
It will be shown later that in the limit of isotropic ($\theta_0\rightarrow \pi$) rotational diffusion this formula yields the well known result VIII.105 from \cite{Ab61}.

\section{Arbitrary orientation of the cone axis relative the magnetic field}
In the general case of arbitrary tilted cone axis relative the magnetic field we need two frames (see Sec.2 and Sec.3). The laboratory fixed frame has the $z$ axis directed along the magnetic field while the dashed (cone-related) frame has the $z'$ axis directed along the cone axis. The angle between the cone axis and the magnetic field is $\psi$. In (\ref{eq13}) we carry out internal averaging over the rotation in the cone in the dashed (cone-related) frame. That is why we need the transformation of the $F^{(i)}\left(\Omega\right)$ given by (\ref{eq14}) in the laboratory fixed frame into those in the dashed (cone-related) frame.

As is well known a rotation of one frame relative the other is most conveniently described by Euler angles. We choose the angle $\psi$ as the first Euler angle ($0 \leq \psi \leq \pi$). We denote two other Euler angles as $\lambda$ (that between the $x$-axis and the so-called $N$-line (node-line) $0 \leq \lambda \leq 2\pi$) and $\omega$ (that between the $N$-line and the $x'$-axis $0 \leq \omega \leq 2\pi$). The formula for transformation of the generalized spherical harmonics at transition from the frame $\{\phi, \theta\}$ to that $\{\phi ', \theta '\}$ obtained by rotation of the $z$ axis by the angle $\psi$ is \cite{Jef65}, \cite{Aks86}
\begin{equation}
\label{eq30} P_2^{(q)}(\cos \theta)\exp (i q \phi) =\sum_{s=-n}^{n} R_{2,q}^{(s)}(\psi) P_2^{(s)}(\cos \theta')\exp \left[i s \phi'+ i q\lambda + i s \omega\right]
\end{equation}
where
\[
R_{2,q}^{(s)}(\psi)=\sum_{r=max(0,-q-s)}^{min(2-q,2-s)} (-i)^{4-2r-q-s}\times
\]
\begin{equation}
\label{eq31} \frac{(2+q)!(2-s)!}{r!(2-q-r)!(2-s-r)!(q+s+r)!}\left(\cos \frac{\psi}{2}\right)^{q+s+2r} \left(\sin \frac{\psi}{2}\right)^{4-q-s-2r}
\end{equation}
One can see that if the distribution function $f(\psi, \lambda, \omega)$ characterizing a system of interest depends only on the angle $\psi$, i.e., $f(\psi, \lambda, \omega)\equiv f(\psi)$ then at overall averaging
\begin{equation}
\label{eq32} <...>_{overall}=\frac{1}{8\pi^2}\int \limits_{0}^{\pi} d\psi \int \limits_{0}^{2\pi} d\lambda \int \limits_{0}^{2\pi} d\omega \ \sin \psi\ f(\psi, \lambda, \omega)...
\end{equation}
we have the factors
\begin{equation}
\label{eq33} \int \limits_{0}^{2\pi} d\lambda\ \exp \left[i(q-q^{\prime})\lambda\right] =2\pi \delta_{qq^{\prime}}
\end{equation}
\begin{equation}
\label{eq34} \int \limits_{0}^{2\pi} d\omega\ \exp \left[i(s-s^{\prime})\omega\right] =2\pi \delta_{ss^{\prime}}
\end{equation}
It is namely the identity (\ref{eq33}) that provides the applicability of the formula (\ref{eq12}). The latter is crucial for the validity of the BPP-Solomon scheme \cite{Ab61}. The overall averaging takes the form
\begin{equation}
\label{eq35} <...>_{overall}=\frac{1}{2} \int \limits_{0}^{\pi} d\psi\ \sin \psi\ f(\psi )...
\end{equation}
and total (overall $+$ internal) averaging is given by (\ref{eq11}) in this case. The function $f(\psi )$ must be normalized so that
\begin{equation}
\label{eq36} \frac{1}{2} \int \limits_{0}^{\pi} d\psi\ \sin \psi\ f(\psi )=1
\end{equation}

We denote
\[
h_{(0)}^{(2)}=\int \limits_{0}^{\pi} d\psi\ \sin \psi f(\psi)\left(\cos \frac{\psi}{2}\right)^4\left(\sin \frac{\psi}{2}\right)^4
\]
\[
h_{(1)}^{(2)}=\int \limits_{0}^{\pi} d\psi\ \sin \psi f(\psi)\left(\cos \frac{\psi}{2}\right)^2\left(\sin \frac{\psi}{2}\right)^2
\left[\left(\cos \frac{\psi}{2}\right)^4+\left(\sin \frac{\psi}{2}\right)^4\right]
\]
\[
h_{(2)}^{(2)}=\int \limits_{0}^{\pi} d\psi\ \sin \psi f(\psi)
\left[\left(\cos \frac{\psi}{2}\right)^8+\left(\sin \frac{\psi}{2}\right)^8\right]
\]
\[
h_{(0)}^{(1)}=\int \limits_{0}^{\pi} d\psi\ \sin^3\psi \cos^2\psi f(\psi)
\]
\[
h_{(1)}^{(1)}=\int \limits_{0}^{\pi} d\psi\ \sin \psi f(\psi)\Biggl\lbrace\frac{1}{4}\sin^4\psi-\cos^2\psi\sin^2\psi+
\]
\[
\cos^2\psi \left[\left(\sin\frac{\psi}{2}\right)^4+\left(\cos\frac{\psi}{2}\right)^4\right]\Biggr\rbrace
\]
\[
h_{(2)}^{(1)}=\int \limits_{0}^{\pi} d\psi\ \sin^3\psi f(\psi)\left[\left(\sin\frac{\psi}{2}\right)^4+\left(\cos\frac{\psi}{2}\right)^4\right]
\]
\[
h_{(0)}^{(0)}=\int \limits_{0}^{\pi} d\psi\ \sin \psi f(\psi)\left(\cos^2\psi-\frac{1}{2}\sin^2\psi\right)^2
\]
\[
h_{(1)}^{(0)}=h_{(0)}^{(1)}
\]
\begin{equation}
\label{eq37} h_{(2)}^{(0)}=\int \limits_{0}^{\pi} d\psi\ \sin^5 \psi f(\psi)
\end{equation}
After lengthy but straightforward calculations we obtain
\begin{equation}
\label{eq38} J^{(2)}(\omega)=2h_{(0)}^{(2)}j^{(0)}(\omega)+8h_{(1)}^{(2)}j^{(1)}(\omega)+\frac{1}{2}h_{(2)}^{(2)}j^{(2)}(\omega)
\end{equation}
\begin{equation}
\label{eq39} J^{(1)}(\omega)=\frac{1}{2}\left[\frac{1}{4}h_{(0)}^{(1)}j^{(0)}(\omega)+h_{(1)}^{(1)}j^{(1)}(\omega)+\frac{1}{4}h_{(2)}^{(1)}j^{(2)}(\omega)\right]
\end{equation}
\begin{equation}
\label{eq40} J^{(0)}(\omega)=\frac{1}{2}h_{(0)}^{(0)}j^{(0)}(\omega)+9h_{(1)}^{(0)}j^{(1)}(\omega)+\frac{45}{16}h_{(2)}^{(0)}j^{(2)}(\omega)
\end{equation}
where $j^{(0)}(\omega)$, $j^{(1)}(\omega)$ and $j^{(2)}(\omega)$ are given by (\ref{eq24})-(\ref{eq26}). In designations (\ref{eq27}) the explicit form of the basic spectral densities $j^{(0)}(\omega)$, $j^{(1)}(\omega)$ and $j^{(2)}(\omega)$ is
\begin{equation}
\label{eq41} j^{(0)}(\omega)=\frac{2\tau}{b^6(1-\mu_0)}\sum_{n=1}^{\infty}\frac{1}{H_n^{(0)}\nu_n^0(\nu_n^0+1)}\left\lbrace K_n^0\right\rbrace^2
\left [1+ \frac{\left (\omega \tau \right )^{2}}{\left[\nu_n^0(\nu_n^0+1)\right]^{2}}\right ]^{-1}
\end{equation}
\[
j^{(1)}(\omega)=\frac{2\tau(1+\mu_0)}{b^6}\sum_{n=1}^{\infty}\frac{(\nu_n^1+3)^{-2}(\nu_n^1-2)^{-2}}{H_n^{(1)}\nu_n^1(\nu_n^1+1)}
\times
\]
\begin{equation}
\label{eq42} \left\lbrace \left[(\nu_n^1+3)\mu_0^2-1 \right]P_{\nu_{n}^1}^{(1)}\left(\mu_0\right)-
\nu_n^1 \mu_0 P_{\nu_{n}^1+1}^{(1)}\left(\mu_0\right)\right\rbrace^2
\left [1+\frac{\left (\omega \tau \right )^{2}}{\left[\nu_n^1(\nu_n^1+1)\right]^{2}}\right ]^{-1}
\end{equation}
\[
j^{(2)}(\omega)=\frac{2\tau(1-\mu_0)^2(1+\mu_0)^3}{b^6}\sum_{n=1}^{\infty}\frac{(\nu_n^2+3)^{-2}(\nu_n^2-2)^{-2}}{H_n^{(2)}\nu_n^2(\nu_n^2+1)}
\times
\]
\begin{equation}
\label{eq43}
\left\lbrace P_{\nu_{n}^2}^{(3)}\left(\mu_0\right)\right\rbrace^2\left [1+\frac{\left (\omega \tau \right )^{2}}{\left[\nu_n^2(\nu_n^2+1)\right]^{2}}\right ]^{-1}
\end{equation}

For practical application of the theory we consider three cases.

1. For the particular case of random isotropic distribution (unweighted average) of cone axes relative the magnetic field we have $f(\psi)=1$. In this case we obtain: $h_{(0)}^{(2)}=1/15$; $h_{(1)}^{(2)}=1/5$; $h_{(2)}^{(2)}=4/5$; $h_{(0)}^{(1)}=4/15$; $h_{(1)}^{(1)}=8/15$; $h_{(2)}^{(1)}=4/5$; $h_{(0)}^{(0)}=2/5$; $h_{(1)}^{(0)}=4/15$; $h_{(2)}^{(0)}=16/15$.

2. As an example of the case for the cone axes to be predominantly oriented along or opposite the magnetic field we consider the model function $f(\psi)=3\cos^2 \psi$. In this case we obtain: $h_{(0)}^{(2)}=1/35$; $h_{(1)}^{(2)}=1/7$; $h_{(2)}^{(2)}=44/35$; $h_{(0)}^{(1)}=12/35$; $h_{(1)}^{(1)}=28/35$; $h_{(2)}^{(1)}=20/35$; $h_{(0)}^{(0)}=22/35$; $h_{(1)}^{(0)}=12/35$; $h_{(2)}^{(0)}=16/35$.\\

3. As an example of the case for the cone axes to be predominantly oriented transverse to the magnetic field we consider the model function $f(\psi)=3/2\sin^2 \psi$. In this case we obtain: $h_{(0)}^{(2)}=3/35$; $h_{(1)}^{(2)}=8/35$; $h_{(2)}^{(2)}=20/35$; $h_{(0)}^{(1)}=8/35$; $h_{(1)}^{(1)}=14/35$; $h_{(2)}^{(1)}=32/35$; $h_{(0)}^{(0)}=10/35$; $h_{(1)}^{(0)}=8/35$; $h_{(2)}^{(0)}=48/35$.\\

Making use of (\ref{eq41})-(\ref{eq43}) and explicit values of $h_{(i)}^{(q)}$ for these three cases enables us to calculate the spectral densities (\ref{eq38})-(\ref{eq40}) to be inserted in BPP-Solomon formulas (\ref{eq9})-(\ref{eq10}). We do not write out explicitly the corresponding expressions for the spin-lattice relaxation rate to save room. However in the next Sec. we plot the corresponding dependencies of spin-lattice relaxation rate for nuclear pair with non-identical spins for all three cases.

\section{Results and discussion}
For $^{15}N - H$ nuclear pair of non-identical spins, we identify $S$ with $^{15}N$ nucleus and $I$ with $H$ one. Thus $\gamma_S=-2712\  rad\ s^{-1}\ Gauss^{-1}$ and $\gamma_I=26753\ rad\ s^{-1}\ Gauss^{-1}$ so that $\gamma_S/\gamma_I=-0.101372$. For
$^{13}C - H$ nuclear pair $\gamma_S=6728\  rad\ s^{-1}\ Gauss^{-1}$ so that $\gamma_S/\gamma_I=0.251486$.
The righthand side in the formula (\ref{eq28}) depends on the parameters characterizing the nuclear pair (namely on the gyromagnetic ratios $\gamma_S$ and $\gamma_I$ of our pair of non-identical spins). To plot the spin-lattice relaxation rate with the help of (\ref{eq28}) one has to choose the particular nuclear pair explicitly. That is why to be specific we choose the $^{15}N - H$ nuclear pair of non-identical spins. It should be mentioned that both (\ref{eq28}) and (\ref{eq29}) have uncertainties at:
$\theta_0=\pi/4$ because $\nu_1^1=2.0000$; $\theta_0=\pi/2$  because $\nu_1^2=2.0000$; $\theta_0=3\pi/4$ because $\nu_2^1=2.0000$; $\theta_0=175^\circ$ because $\nu_1^2=2.0000$. However these uncertainties are isolated, and can be safely ignored. 

In Fig. 1 the spin-lattice relaxation rate for identical spins obtained with the help of (\ref{eq29}) for the case of cone axes directed along the magnetic field is depicted as a function of the cone half-width $\theta_0$ at different values of Larmor frequency. The value $\omega_{L} \tau =0.1$ for the upper curve is within the range of validity of the extreme narrowing limit $\omega_{L} \tau << 1 $.
From this Fig. one can see that for the case of isotropic ($\theta_0=\pi$) rotational diffusion in the limit of extreme narrowing the corresponding curve tends to the value $0.333...$. Thus the formula (\ref{eq29}) yields
\begin{equation}
\label{eq44} \lim_{\theta_0\to\pi}\left( \frac{b^6}{\tau\gamma^4 \hbar^2 I(I+1)}\right)
\left(\frac{1}{T_{1}}\right)_{rotat}\left| {\begin{array}{l}
  \\
\alpha=1\\
 \end{array}}\right. =\frac{1}{3}
\end{equation}
that taking into account (\ref{eq27}) coincides with the well known formula VIII.106 from \cite{Ab61}
\[
\left(\frac{1}{T_{1}}\right)_{rotat}=\frac{2\gamma^4\hbar^2 }{b^6}I(I+1)\frac{4\pi\eta a^3}{3k_BT}
\]

In Fig.2 the spin-lattice relaxation rate for non-identical spin pair $^{15}N - H$ obtained with the help of (\ref{eq38})-(\ref{eq40})  for the case of random isotropic distribution (unweighted average $f(\psi)=1$ taking place in powders) of the cone axes relative the external magnetic field is depicted as a function of the cone half-width $\theta_0$ at different values of Larmor frequency.
The reduced curves $\left[ b^6/\left(\tau\gamma_I^2\gamma_S^2 \hbar^2 S(S+1)\right)\right]\left(1/{T_{1}^{II}}\right)_{rotat}$
for $^{13}C - H$ nuclear pair are found to be very similar to those for $^{15}N - H$ one. For this reason data for $^{13}C - H$ nuclear pair are not presented to save room.

In Fig. 3 the dependence of the spin-lattice relaxation rate on the cone half-width $\theta_0$ for the case of non-identical spin pair $^{15}N - H$ is depicted for different distributions of the cone axis relative the external magnetic field: for the case of cone axes directed along the magnetic field ($f(\psi)=\delta (\psi)$); for the case of the cone axes to be predominantly oriented along or opposite the magnetic field ($f(\psi)=3\cos^2 \psi$ that may be relevant for, e.g., liquid crystals); for the case of random isotropic distribution (unweighted average $f(\psi)=1$ taking place in powders); for the case of the cone axes to be predominantly oriented transverse to the magnetic field ($f(\psi)=3/2\sin^2 \psi$ that may be relevant for, e.g., liquid crystals).

As is well known in liquids (for which isotropic rotation is relevant) $T_1$ ordinarily decreases with increasing viscosity, in some cases reaching a minimum value after which it increases with further increase in viscosity  \cite{Ab61}.
The variation of viscosity is caused by temperature $T$. For ordinary isotropic rotation correlation times are functions of temperature $\tau_n =\tau \left(\eta (T)\right)[n(n+1)]^{-1}$ (see VIII.97 in \cite{Ab61}) and the spin-lattice relaxation rate $(1/T_{1})_{rotat} \left( \tau (T)\right)$ can have a maximum as a result. In the present model we have one more option. Substitution of (\ref{eq27}) into (\ref{eq23}) yields
\begin{equation}
\label{eq45} \tau^{(m)}_n=\frac{\tau \left(\eta (T)\right)}{\nu_n^m(\nu_n^m+1)}
\end{equation}
where $m=0,1,2$. Thus correlation times are functions of temperature and of the cone half-width $\theta_0$ (via the values $\nu_n^m$).  The results obtained testify that variation of each of these parameters ($T$ or $\theta_0$) can produce a maximum in the corresponding dependence of the spin-lattice relaxation rate. The dependence in Fig.1 exhibits a maximum at $\theta_0 \approx 100^\circ$. Fig.2 and Fig. 3 show similar maximums for non-identical spin pair at $\theta_0 \approx \pi/2$.
These maximums are similar to those at $\theta_0 \approx \pi/2$ given by the model-free approach.

Indeed let us consider the most typical for practice case of the model-free approach when the overall motion of a macromolecule is considerably slower than the internal motion. In this case the expression of the model-free approach for the relationship of the spin-lattice relaxation rate with the order parameter $S^2$ and effective correlation time $\tau_e$ is given by  equation (37) from \cite{Lip82}
\begin{equation}
\label{eq46} \frac{1}{T_1}=aS^2+b\tau_e\left(1-S^2\right)
\end{equation}
where $a$ and $b$ are constants independent on spatial configuration accessible for internuclear vector. For wobbling in a cone the relationship between the order parameter of the model-free approach and the cone half-width $\theta_0$  is given by equation (A3) from \cite{Lip82}
\begin{equation}
\label{eq47} S=\frac{1}{2}\cos \theta_0 (1+\cos \theta_0)
\end{equation}
while that for the effective correlation time is given by equation (A4) from \cite{Lip82}
\[
\tau_e=\frac{1}{D_w\left(1-S^2\right)}\Biggl\{\cos^2\theta_0\left(1+\cos \theta_0\right)^2
\bigl\{\ln\left[(1+\cos \theta_0)/2\right]+
\]
\[
(1-\cos \theta_0)/2\bigr\}/[2(\cos \theta_0-1)]+(1-\cos \theta_0)(6+8\cos \theta_0-
\]
\begin{equation}
\label{eq48} \cos^2 \theta_0-12\cos^3 \theta_0-7\cos^4 \theta_0)/24\Biggr\}
\end{equation}
where $D_w$ is the diffusion coefficient. We denote
\begin{equation}
\label{eq49} c=\frac{b}{aD_w}
\end{equation}
Then we obtain the dependence of the spin-lattice relaxation rate on the cone half-width $\theta_0$
\[
\frac{1}{a}\frac{1}{T_1}=\frac{1}{4}\cos^2 \theta_0 (1+\cos \theta_0)^2+c\Biggl\{\cos^2\theta_0\left(1+\cos \theta_0\right)^2
\bigl\{\ln\left[(1+\cos \theta_0)/2\right]+
\]
\[
(1-\cos \theta_0)/2\bigr\}/[2(\cos \theta_0-1)]+(1-\cos \theta_0)(6+8\cos \theta_0-
\]
\begin{equation}
\label{eq50} \cos^2 \theta_0-12\cos^3 \theta_0-7\cos^4 \theta_0)/24\Biggr\}
\end{equation}
The dependence of this spin-lattice relaxation rate on the cone half-width $\theta_0$ is depicted in Fig.6 at several values of the parameter $c$. One can see close qualitative similarity of the curves with our results.

The BPP-Solomon scheme requires essential extent the isotropy (but not total isotropy) for its validity. In Sec.6 we show that in application to rotational diffusion in a cone it remains valid for systems with a distribution of cone axes depending only on the tilt  relative the magnetic field but otherwise being isotropic. This residual isotropy provides the requirement (\ref{eq33}) that is
necessary for the absence of cross-correlational functions (\ref{eq12}). The latter in turn is crucial for the validity of of the BPP-Solomon scheme. We consider the aforesaid a so important issue that would like to reiterate it in other words with complete definiteness. We develop the theory for the general case of arbitrary orientation of the cone axis relative the magnetic field (laboratory fixed frame). We show that when the cone axis is tilted at an arbitrary angle $\psi$ to the magnetic field but otherwise is oriented isotropically then at overall averaging the crucial requirement for the validity of the Bloemberger, Purcell, Pound - Solomon scheme (\ref{eq12}) (that of the absence of cross-correlational functions with $q \not= q'$) is retained. Thus we explicitly prove the consistency of combination of the textbook formulas for the BPP-Solomon scheme with rotational motion in a cone.

Practical applications of our results depend on the choice of a distribution function for overall averaging over all orientations of the cone axis with respect to the laboratory fixed frame (over the angle $\psi$). This distribution function $f(\psi)$ is a characteristic of the system of interest. Under the assumption of isotropic random orientation of cone axes relative the laboratory fixed frame the results obtained can be applied to powders. We consider this case of unweighted average over all orientations of the cone axis with respect to the laboratory fixed frame ($f(\psi)=1$) in Sec.6 and provide corresponding data for the spin-lattice relaxation rate in Fig.3. Also in Sec.6 we consider a model example of predominant orientation of the cone axis along or opposite the magnetic field ($f(\psi)=3\cos^2 \psi$) and that of their predominant orientation transverse to the magnetic field  ($f(\psi)=3/2\sin^2 \psi$). The results for these cases may be relevant for, e.g., liquid crystals. For both of these cases we also provide corresponding data for the spin-lattice relaxation rate in Fig.3. The particular case of the cone axis directed along the magnetic field ($f(\psi)=\delta (\psi)$ where $\delta (x)$ is a Dirac delta-function) is of little practical significance. However this case provides direct comparison of the limiting case of our formulas with the textbook formulas from \cite{Ab61}. Thus it serves as a test for the validity of our approach from the theoretical side. 

Fig.3 shows that the results for practically relevant cases of specific distributions of cone axes relative the magnetic field are qualitatively similar to those for the model case of cone axes directed along the magnetic field. However there is some quantitative difference. From Fig.3 one can see that
\[
 \left(\frac{1}{T_1}\right)_{f(\psi)=\delta (\psi)} < \left(\frac{1}{T_1}\right)_{f(\psi)=3\cos^2 \psi} < \left(\frac{1}{T_1}\right)_{f(\psi)=1} < \left(\frac{1}{T_1}\right)_{f(\psi)=3/2\sin^2 \psi}
\]
Thus the more is the contribution of the orientations of the cone axes in the system transverse to the magnetic field (i.e., the more is the  probability of such orientations) the more efficiently spin-lattice relaxation proceeds. This dependence is weak but well-defined.
To prove it explicitly we plot in Fig.4 the the dependence of the spin-lattice relaxation rate on the tilt angle of the cone axis relative the magnetic field for the cone with the half-width $\theta_0 = 55^\circ$ as an example. This case corresponds to the distribution function $f(\psi)=\delta (\psi-\psi_0)\left[\cos \left(\psi_0/2\right)\right]^{-1}$ (see Appendix B for technical details). One can see that when the cone axis is transverse the magnetic field the spin-lattice relaxation proceeds most efficiently. In Fig.5 such dependence is plotted for different values of the cone half-width $\theta_0 $.

The rigorous quantum-mechanical treatment presented in the main text of the paper does not allow simple physical interpretation of the results obtained. To attain the latter objective we invoke to the results of the classical analogy presented in Appendix C. Such simplified approach does not yield precise results but enables us to get more clear physical perception of the phenomenon under consideration. We conclude that the essence of the maximum in the dependence of the spin-lattice relaxation rate on the cone half-width $\theta_0$ originates from two factors:\\
1. the angular dependence of the spin-lattice relaxation rate
\[
\frac{1}{T_1} \propto \cos^2\theta\sin^2\theta
\]
and\\
2. the averaging of the spin-lattice relaxation rate over the rotational diffusion in the cone with half-width $\theta_0$
\[
\left <\frac{1}{T_1}\right >_{cone}\propto \frac{1}{1-\cos \theta_0}\int \limits_{0}^{\theta_0} d\theta\ \sin \theta\ \frac{1}{T_1}
\]
Combination of these two factors leads to the required non-monotonic behavior (see Fig.7). Regretfully further simplification of the physical picture underlying the phenomenon seems hardly possible. It is hampered namely by the fact that the phenomenon can not be attributed to a single reason but originates from the combination of two of them.

Thus the model yields an unexpected result: too much spatial freedom is not good for spin-lattice relaxation rate. This result is obtained from rigorous quantum-mechanical treatment, is loosely supported by its rough classical analogy and besides is well corroborated by the model-free approach. The restriction of free isotropic motion ($\theta_0=\pi$) to smaller cone half-widths leads to the increase in spin-lattice relaxation rate with a maximum in the region of $\theta_0^{max}\approx 90^\circ \div 100^\circ$. Only further restriction of the cone semi-angle to the values $\theta_0 < \theta_0^{max}$ leads at last to efficient decrease of the spin-lattice relaxation rate. We can reformulate the above mentioned physical interpretation of this phenomenon in other words. The maximum arises at $\theta_0 \approx\theta_0^{max}$ because in this case the internuclear vector has maximal probability to be transverse the external magnetic field. As one can see from Fig.4 and Fig.5 such orientation maximizes the spin-lattice relaxation rate. Increase of $\theta_0$ from $\theta_0^{max}$ to the limit of free isotropic motion ($\theta_0=\pi$) decreases the probability for internuclear vector to be transverse the external magnetic field and thus leads to the decrease of the spin lattice relaxation rate. Taking into account that wobbling in a cone is the exactly tractable model we can not attribute this result to any approximations. Also the model captures the effect of restriction for the motion in an adequate and non-trivial manner. Thus the phenomenon can not be model dependent. We conclude that the phenomenon should be of general character, i.e., any correct model of restricted rotational diffusion should exhibit a non-monotonic dependence of the spin-lattice relaxation rate on the accessible space volume for the internuclear vector. The maximum for this dependence should take place for those configurations for which the probability of the inter-nuclear vector to be transverse the external magnetic field is maximal.

This result can manifest itself in experiments where addition of some substance creates efficient steric hindrance to rotating internuclear vector of the reporting nuclear spin pair. Its accessible space volume initially (without this substance) ought to be large enough so that one can approximate it by the limit of free isotropic motion $\theta_0 \rightarrow \pi$. Also one should take special concern that the additional substance does not alter the micro-viscosity for the environment of the reporting spin pair appreciably. It can be achieved if the size of the molecules for the added substance is not small but rather commensurable with those containing the reporting nuclear spin pair. Only in this case one can obtain the required condition that variation of the concentration $C$ of the added substance touches upon primarily $\nu_n^m (C)$ via the accessible space (cone half-width) while micro-viscosity remains mainly constant $\eta (C) \approx const$. Otherwise we are in danger that variation of the correlation times
\[
\tau^{(m)}_n=\frac{\tau \left(\eta (C)\right)}{\nu_n^m (C)(\nu_n^m (C)+1)}
\]
is attributed namely to $\eta (C)$ rather than to $\nu_n^m (C)$. In the latter case one actually probes the dependence $1/T_1$  vs. micro-viscosity $\eta (C)$ with non-monotonic behavior of quite the same nature as that for the usual dependence $1/T_1$  vs. temperature $T$ stipulated by the dependence $\eta (T)$. Provided all precautions are taken into account and the above condition is satisfied one may hope that the dependence $1/T_1$  vs. cone half-width $\theta_0$ can be probed in the experiment.
Then basing on the results obtained in the present paper (Fig.1, Fig.2 and Fig.3) one can expect that the increase of the concentration of the additional substance should lead to the increase of the spin-lattice relaxation rate up to a maximum with further decrease. The results of our rigorous treatment will be more accurate for the description of such experiments than those of the approximate approach \cite{Lip82}, \cite{Lip811}. Indeed Fig.6 shows that the latter predicts the maximum strictly at one value of the cone half-width, namely at $\theta_0^{max} \approx 85^\circ$. On the other hand Fig.1, Fig.2 and Fig.3 show that within the framework of the rigorous treatment the position of the maximum is a variable value from the range $80^\circ < \theta_0^{max}  < 180^\circ$. It depends on the type of the nuclear spin pair (homonuclear or heteronuclear), on the Larmor frequency or more exactly on the value of $\omega_L \tau$, etc. The dependence of the maximum position on the type of the labeled atom ($^{15}N $ or $^{13}C $) is found to be very small. Thus we anticipate that the results of the present approach leave much more freedom for describing experimental data.

We conclude that wobbling in a cone is the exactly tractable (i.e., amenable to full analytic treatment within the range of validity of BPP-Solomon scheme) model for description of the nuclear spin-lattice relaxation rate. This fact makes it a unique one among all other models for restricted rotational diffusion.

Acknowledgements.  The author is grateful to Dr. Yu.F. Zuev and R.H. Kurbanov for
helpful discussions. The work was supported by the grant from RFBR and
the programme "Molecular and Cellular Biology" of RAS.

\section{Appendix A}
Here we present two known mathematical formulas 1.12.1.12 and 1.12.1.9 for the the associated Legendre function $P_{\nu}^{\mu}\left(x\right)$ from the table of integrals \cite{Pru03}. The formula 1.12.1.12 is
\[
 \int  dx\  x(1-x^2)^{\pm \mu/2}P_{\nu}^{\mu}\left(x\right)=\frac{(1-x^2)^{\pm \mu/2}}{(\nu \pm \mu +2)(\nu \mp \mu -1)}\times
\]
\[
\left \{ \left[(\nu \pm \mu +2)x^2-1\right]P_{\nu}^{\mu}\left(x\right)+(\mu - \nu -1)xP_{\nu+1}^{\mu}\left(x\right)\right\}
\]
The formula 1.12.1.9 is
\[
 \int  dx\  (1-x^2)^{\mu/2}P_{\nu}^{\mu}\left(x\right)=\frac{(1-x^2)^{(\mu+1)/2}}{(\nu - \mu)(\nu +\mu +1)}P_{\nu}^{\mu+1}\left(x\right)
\]

\section{Appendix B}
The polar angle $\psi$ at operations with the distribution function $f(\psi)$ imposes some peculiarities in treating the case of Dirac $\delta-$function
\[
f(\psi)=\frac{1}{\cos \left(\psi_0/2\right)}\delta (\psi-\psi_0)
\]
We stress that the case of the cone angle oriented along the magnetic field $f(\psi)=\delta (\psi)$ considered in Sec. 4- Sec.6 is a particular case of this distribution function corresponding to the value $\psi_0=0$. First let us prove the normalization requirement (\ref{eq36}). We have (taking into account that $\delta (2z)=\delta (z)/2$)
\[
\frac{1}{2} \int \limits_{0}^{\pi} d\psi\ \sin \psi\ \delta (\psi-\psi_0)=2\int \limits_{0}^{\pi} d\left(\frac{\psi}{2}\right) \sin \frac{\psi}{2} \cos \frac{\psi}{2} \delta \left(2\frac{\psi-\psi_0}{2}\right)=
\]
\[
\int \limits_{0}^{\pi/2} dx\ \sin x \cos x\ \delta \left(x-\frac{\psi_0}{2}\right)=-\int \limits_{0}^{\pi/2} d(\cos x)\ \cos x\ \delta \left(x-\frac{\psi_0}{2}\right)=
\]
\[
\int \limits_{0}^{1} dy\ y\ \delta \left(\arccos y - \frac{\psi_0}{2}\right)=\int \limits_{0}^{1} dy\ y\ \delta \left(y - \cos \frac{\psi_0}{2}\right)= \cos \frac{\psi_0}{2}
\]
This calculation serves as a model for operations at calculating the values of $h_{(m)}^{(n)}$ in (\ref{eq37}) with the distribution function $f(\psi)=\delta (\psi-\psi_0)\left[\cos \left(\psi_0/2\right)\right]^{-1}$. The general rule takes the form
\[
\frac{1}{\cos \left(\psi_0/2\right)}\int \limits_{0}^{\pi} d\psi\ \sin \psi\ \delta (\psi-\psi_0)q(\psi) =2q(\psi_0)
\]
where $q(\psi)$ is an arbitrary function of $\psi$.

At $\psi_0=0$ we obtain $h_{(2)}^{(2)}=2$, $h_{(1)}^{(1)}=2$ and $h_{(0)}^{(0)}=2$ while $h_{(m)}^{(n)}=0$ at $m\not=n$. Substitution of these values into (\ref{eq38})-(\ref{eq40}) yields $J^{(2)}(\omega )=j^{(2)}(\omega )$,  $J^{(1)}(\omega )=j^{(1)}(\omega )$ and  $J^{(0)}(\omega )=j^{(0)}(\omega )$ as it must be.

\section{Appendix C}
The rigorous quantum-mechanical treatment described in the text of the present paper does not enable us to gain insight in physical interpretation of the results obtained. To attain the latter objective we present here its classical analogy that is much more simple and not at all precise but nevertheless rather illustrative. Let us consider the simplest case of two identical spins (nuclei) with magnetic moments $\overline \mu$ which we assume for simplicity to be oriented along the external magnetic field $\overline H_0$ directed along the axis $z$ of the spherical frame so that $\overline \mu=\mu \overline e_z$. The internuclear vector (with the absolute value $r$) is tilted by the angle $\theta$ relative the external magnetic field. We denote $\overline n$ the unit vector for internuclear one. Then the magnetic field produced by one spin in the location of the other is
\[
\overline h \propto \frac{1}{r^3}\left[\overline \mu - \left(\overline \mu \cdot  \overline n\right) \overline n\right]
\]
The spin-lattice relaxation rate for classical treatment of dipole-dipole interaction is given by the following expression (see, e.g., Appendix E in \cite{Cow97})
\[
\frac{1}{T_1}=\gamma^2 \int \limits_{0}^{\infty} d\tau \cos \left(\omega_L \tau\right)\left[<h_x(0)h_x(\tau)>+<h_y(0)h_y(\tau)>\right]
\]
where $\gamma$ is the gyromagnetic ration and $\omega_L=\gamma H_0$ is the Larmor frequency.

For our diffusion in a cone model we have to average the latter value over the rotational motion in the cone with half-width $\theta_0$. Thus we want to calculate the value
\[
R_1\equiv\left <\frac{1}{T_1}\right >_{cone}\propto \frac{1}{1-\cos \theta_0}\int \limits_{0}^{\theta_0} d\theta\ \sin \theta\ \frac{1}{T_1}
\]
The unit vector $\overline n$ in the spherical frame is
\[
\overline n = \sin \theta \cos \phi \overline e_x + \sin \theta \sin \phi \overline e_y +  \cos \theta \overline e_z
\]
so that
\[
\left(\overline \mu \cdot  \overline n\right) =\mu \left(\overline e_z \cdot  \overline n\right) =\mu \cos \theta
\]
One can see that both $\ h_x\propto \cos\theta\sin\theta\ $ and $\ h_y\propto \cos\theta\sin\theta\ $ so that
\[
\frac{1}{T_1} \propto \cos^2\theta\sin^2\theta
\]
The origin of the latter relationship can be also trivialized as follows. The energy for the  dipole-dipole interaction of two magnetic moments is
\[
U \propto 3\cos^2\theta -1
\]
Then the force of interaction (this notion is quite viable in our classical analogy of the quantum-mechanical treatment) is
\[
F \propto \frac{dU}{d\theta}\propto \cos\theta\sin\theta
\]
The spin-lattice relaxation rate is determined by the correlation function of the force that yields the above relationship.
Thus we obtain that the angular dependence of the spin-lattice relaxation rate on the cone half-width $\theta_0$ is
\[
R_1\propto \frac{1}{1-\cos \theta_0}\int \limits_{0}^{\theta_0} d\theta\ \sin^3 \theta\ \cos^2 \theta
\]
Integration yields the final result
\[
R_1\propto \frac{2}{15}\left(1+ \cos \theta_0 + \cos^2 \theta_0 \right)-\frac{1}{5}\cos^3 \theta_0 \left(1+\cos \theta_0\right)
\]
This dependence is depicted in Fig.6.

\newpage
\clearpage
\begin{figure}
\begin{center}
\includegraphics* [width=\textwidth] {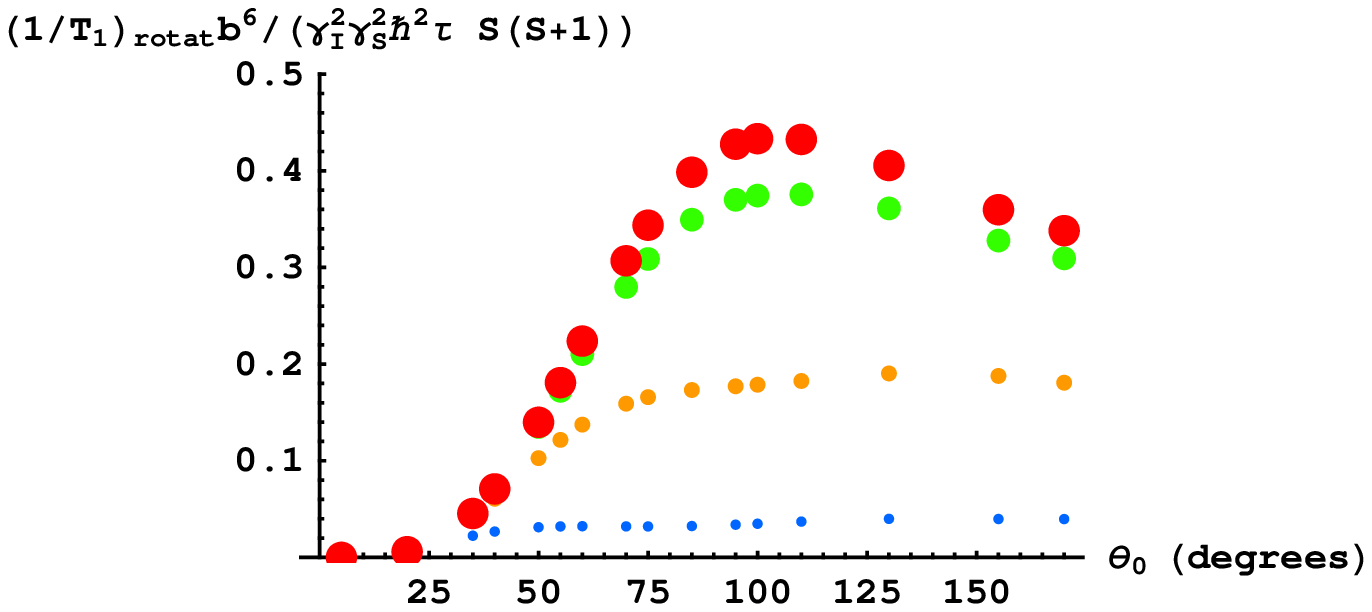}
\end{center}
\caption{Spin-lattice relaxation rate for nuclear pair of identical spins from rotational diffusion for the case of cone axes directed along the magnetic field (eq. (\ref{eq29})) as the function of the cone half-width $\theta_0$ (in degrees) at different values of the Larmor frequency: $\omega_L \tau=10$ (thin dots), $\omega_L \tau=10^{0.5}$, $\omega_L \tau=1$, $\omega_L \tau=0.1$ (thick dots). For the case of free isotropic rotational diffusion (cone half-width $\theta_0 \longrightarrow \pi$) the textbook value $0.333...$ ( i.e., formula VIII.106 from \cite{Ab61}) is obtained as it must be (see (eq. (\ref{eq44}) and text below).}
\label{Fig.1}
\end{figure}

\clearpage
\begin{figure}
\begin{center}
\includegraphics* [width=\textwidth] {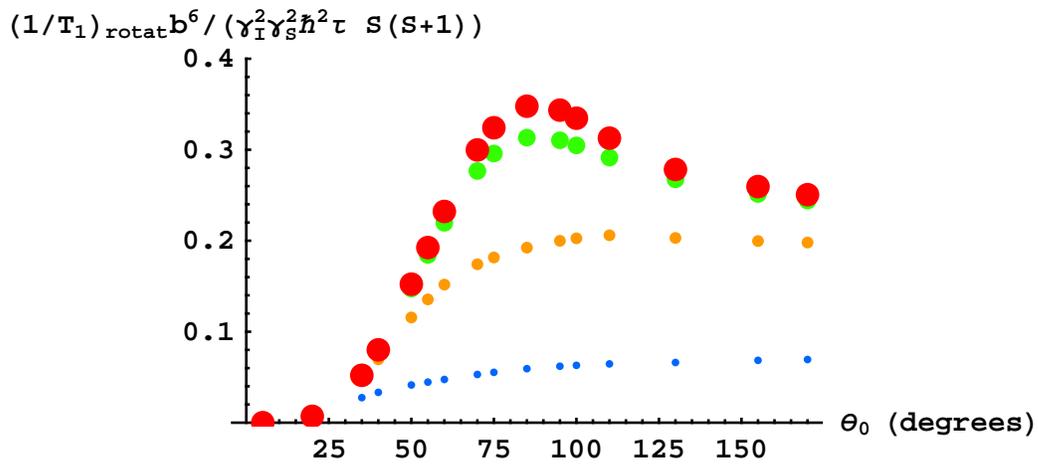}
\end{center}
\caption{Spin-lattice relaxation rate for $^{15}N - H$ nuclear pair of non-identical spins from rotational diffusion for the case of random isotropic distribution (unweighted average $f(\psi)=1$ taking place in powders) of the of the cone axis relative the laboratory fixed frame (external magnetic field) as the function of the cone half-width $\theta_0$ (in degrees) at different values of the Larmor frequency: $\omega_L \tau=10$ (thin dots), $\omega_L \tau=10^{0.5}$, $\omega_L \tau=1$, $\omega_L \tau=0.1$ (thick dots).}
\label{Fig.2}
\end{figure}

\clearpage
\begin{figure}
\begin{center}
\includegraphics* [width=\textwidth] {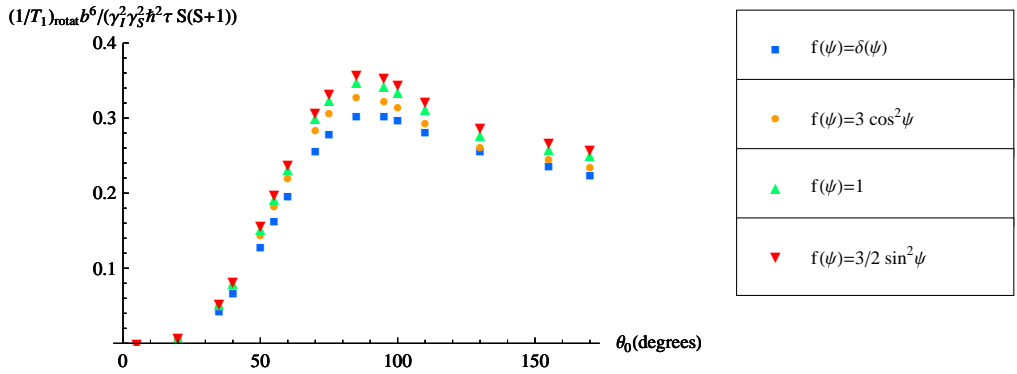}
\end{center}
\caption{Spin-lattice relaxation rate for $^{15}N - H$ nuclear pair of non-identical spins from rotational diffusion  as the function of the cone half-width $\theta_0$ (in degrees) at  $\omega_L \tau=0.1$ for
different distributions of the of the cone axis relative the laboratory fixed frame (external magnetic field): for the case of cone axes directed along the magnetic field ($f(\psi)=\delta (\psi)$); for the case of the cone axes to be predominantly oriented along or opposite the magnetic field ($f(\psi)=3\cos^2 \psi$ that may be relevant for, e.g., liquid crystals); for the case of random isotropic distribution (unweighted average $f(\psi)=1$ taking place in powders); for the case of the cone axes to be predominantly oriented transverse to the magnetic field ($f(\psi)=3/2\sin^2 \psi$ that may be relevant for, e.g., liquid crystals).}
\label{Fig.3}
\end{figure}

\clearpage
\begin{figure}
\begin{center}
\includegraphics* [width=\textwidth] {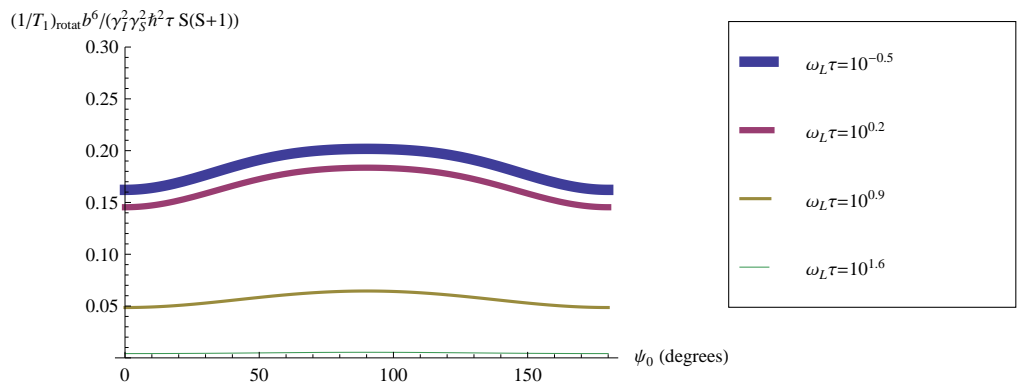}
\end{center}
\caption{Spin-lattice relaxation rate for $^{15}N - H$ nuclear pair of non-identical spins from rotational diffusion in the cone with the half-width $\theta_0=55^\circ$ as the function of the tilt angle $\psi_0$ of the cone axis relative the magnetic field
(corresponding to the distribution function $f(\psi)=\delta (\psi-\psi_0)\left[\cos \left(\psi_0/2\right)\right]^{-1}$)
for
different values of the Larmor frequency: $\omega_L \tau=10^{-0.5}$ (thick line); $\omega_L \tau=10^{0.2}$; $\omega_L \tau=10^{0.9}$ ; $\omega_L \tau=10^{1.6}$ (thin line).}
\label{Fig.4}
\end{figure}

\clearpage
\begin{figure}
\begin{center}
\includegraphics* [width=\textwidth] {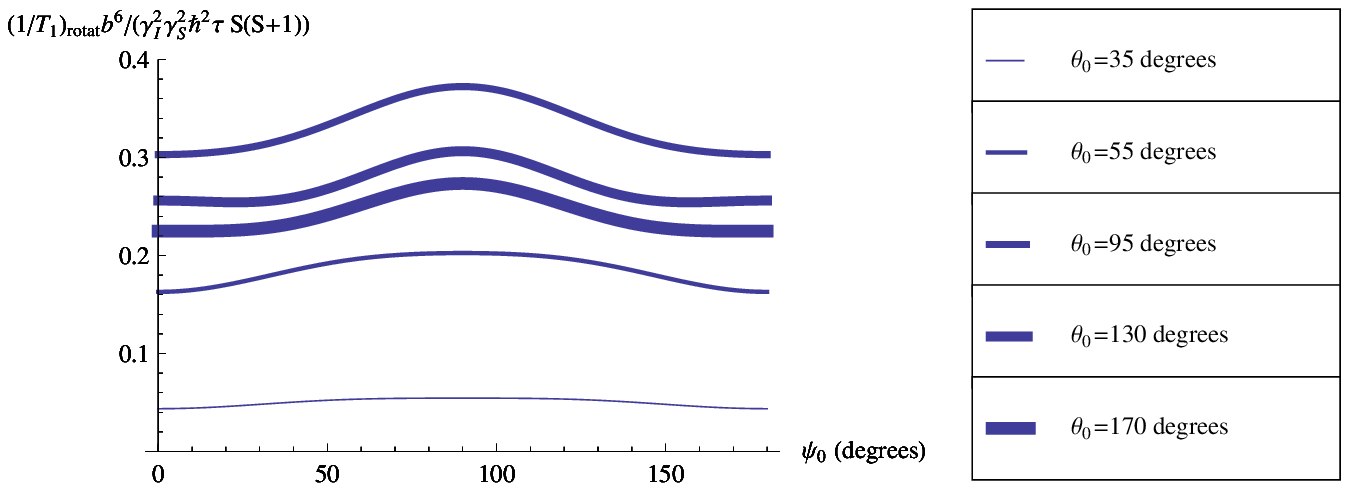}
\end{center}
\caption{Spin-lattice relaxation rate for $^{15}N - H$ nuclear pair of non-identical spins from rotational diffusion in the cone at $\omega_L \tau=0.1$ as the function of the tilt angle $\psi_0$ of the cone axis relative the magnetic field
(corresponding to the distribution function $f(\psi)=\delta (\psi-\psi_0)\left[\cos \left(\psi_0/2\right)\right]^{-1}$)
for
different values of the cone half-width: $\theta_0=35^\circ$ (thin line); $\theta_0=55^\circ$; $\theta_0=95^\circ$; $\theta_0=130^\circ$; $\theta_0=170^\circ$ (thick line).}
\label{Fig.5}
\end{figure}

\clearpage
\begin{figure}
\begin{center}
\includegraphics* [width=\textwidth] {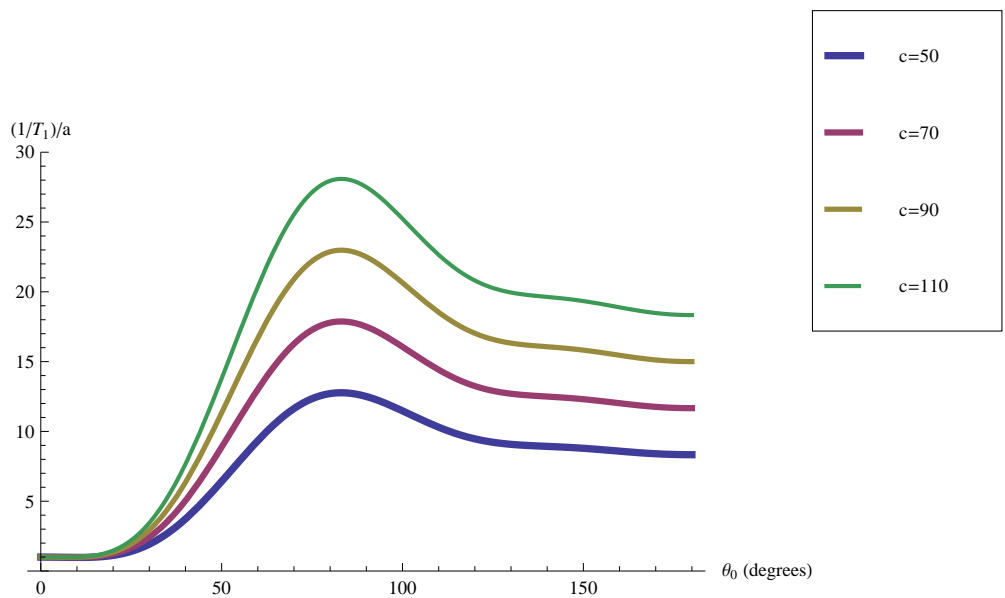}
\end{center}
\caption{Spin-lattice relaxation rate for the model-free approach (eq. (\ref{eq50})) as the function of the cone half-width $\theta_0$ for
different values of the dimensionless parameter $c$ (eq. (\ref{eq49})): $c=50$ (thick line); $c=70$ ; $c=90$ ; $c=110$  (thin line).}
\label{Fig.6}
\end{figure}

\clearpage
\begin{figure}
\begin{center}
\includegraphics* [width=\textwidth] {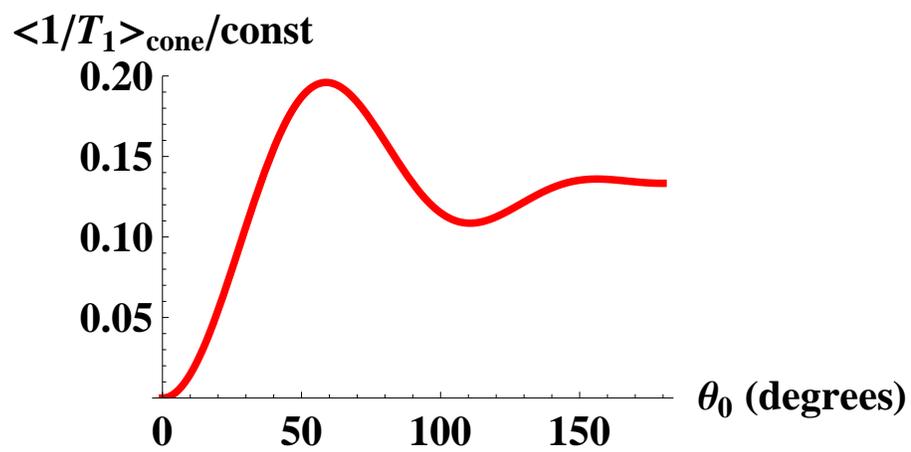}
\end{center}
\caption{Angular dependence of the spin-lattice relaxation rate on the cone half-width $\theta_0$ for classical analogy of the rigorous treatment of diffusion in a cone model (see Appendix C).}
\label{Fig.7}
\end{figure}

\end{document}